# Quantifying and Correlating Rhythm Formants in Speech

Dafydd Gibbon[1] and Peng Li[2]
[1]Bielefeld University Germany
[2]Guangdong University of Finance, China
gibbon@uni-bielefeld.de, 15521323216@163.com

## Abstract

The objective of the present study is exploratory: to introduce and apply a new theory of speech rhythm zones or rhythm formants (R-formants). R-formants are zones of high magnitude frequencies in the low frequency (LF) long-term spectrum (LTS), rather like formants in the short-term spectra of vowels and consonants. After an illustration of the method, an R-formant analysis is made of non-elicited extracts from public speeches. The LF-LTS of three domains, the amplitude modulated (AM) absolute (rectified) signal, the amplitude envelope modulation (AEM) and frequency modulation (FM, F0, 'pitch') of the signal are compared. The first two correlate well, but the third does not correlate consistently with the other two, presumably due to variability of tone, pitch accent and intonation. Consequently, only the LF LTS of the absolute speech signal is used in the empirical analysis. An informal discussion of the relation between R-formant patterns and utterance structure and a selection of pragmatic variables over the same utterances showed some trends for R-formant functionality and thus useful directions for future research.

## 1. Introduction

The study of rhythm as a meaningful property of speech can be grounded in a fairly consensual semiotics of prosody:

> Rhythms in music and language are semiotic events: regularly repeated temporal patterns of human experience in performing and perceiving music, dance and speech, and, more metaphorically, in events of non-human origin such as animal sounds, and in regularly repeated spatial patterns in the visual arts and in the dynamics of natural phenomena.
> (Gibbon and Lin in press)

Two complementary approaches to the study of rhythm in this sense have emerged. The first is a deductive approach which starts with linguistically defined categories such as the syllable, the foot, or consonantal and vocalic features, and which then relates these to correlates in the speech signal by means of annotating the speech signal with these linguistic categories. The second is an inductive approach which starts with the speech signal and induces analyses by means of transformations of the speech signal to reveal rhythmical patterns in both time and frequency domains. After initial detailed discussion of the traditional deductive approach to rhythm analysis, the study focuses on the inductive approach in the form of *Rhythm Formant Theory* (*RFT*, previously: *Rhythm Zone Theory*, *RZT*), which shows a number of further developments in relation to previous work.

The objective of the present study is exploratory: to examine rhythmically relevant formant-like zones of higher magnitude (R-formants) in the low frequency (LF) region of long term spectra (LTS) of the speech signal, using non-elicited extracts from public speeches. LF patterns of three LTS domains are compared:

1. amplitude modulation (AM) of the signal, for example syllabic amplitude patterns in the waveform;
2. amplitude envelope modulation (AEM, a property of AM);
3. frequency modulation (FM, i.e. F0, 'pitch').

The null hypothesis claims significant correlation between each pair of the three spectral domains while the alternative hypothesis claims differences, predicted to depend on divergence between grammar and information structure, or differences in pitch accent, tone and intonation contours. A tentative study of the relation between these domains and a pragmatic variable over the same utterances was also made.

Section 2 places the theory of LF rhythms and R-formants in the context of earlier studies of speech rhythms, and extends the concentration in earlier approaches on the AM and AEM spectra (AMS, AEMS) to the FM LTS. Section 3 outlines the data and methods used, followed by discussion of the results in Section 4. Finally, in Section 5, the outcomes are summarised, conclusions are drawn and the outlook for future work is sketched.

## 2. Rhythm analysis paradigms

### 2.1. Deduction: annotation-based isochrony metrics

Discussions of speech rhythm in phonetics and linguistics during the past fifty years have been mainly deductive in the sense outlined in Section 1. Deductive approaches have been diverse and controversial, and have not always addressed the core feature of standard characterisations of rhythm as alternation or oscillation. Conversational analysis has provided holistic hermeneutic judgments of rhythms in discourse (Couper-Kuhlen 1993) as informal percepts of properties of the speech signal. In the grammar domain, including morphology and phonology, rationalist reconstruction of categories of word and sentence stress placement has modelled rhythms as intuitively motivated abstract structures, with an informal understanding of rhythms as stress patterns (Chomsky and Halle 1968; Liberman and Prince 1977; Kager 1999 and many later studies in generative and post-generative paradigms).

In phonetics, several empirical deductive approaches have also been developed, starting with simple distinctions between mora, syllable and foot timing, and progressing to quantitative scales. The most popular approaches for over half a century have measured the relative isochrony of phonological units such as consonantal and vocalic speech segments or syllables and feet, that is, the degree to which sequences of these units have similar durations (Roach 1982, Jassem et al. 1984; Scott et al. 1985 and many later studies). The simplest of these isochrony measures is standard deviation, and the others are also essentially measures of dispersion around the mean. It is evident that such global dispersion measures are not models of rhythm but heuristic indices of relative evenness of durations.

The most widely used and most successful isochrony measure is the *Pairwise Variability Index* (*PVI*) genre of isochrony measures. It is worth analysing the measures in detail. The raw and normalised *PVI* variants, *rPVI* and *nPVI*, respectively, were introduced by Low et al. (2000). Unlike earlier measures, the *PVI* variants do not measure dispersion from the mean, but average differences between adjacent values in a vector of durations, in order to track and neutralise the effect of speech rate change. The *rPVI* is typically used for consonantal sequences, whose duration is relatively invariant, and the *nPVI* is typically used for vocalic sequences, which tend to vary as a function of changes in speech rate. The *PVI* variants are formulated over a vector $D = (d_1, …, d_n)$ of durations of consonantal, vocalic, syllabic, etc., chunks of the speech signal:

$$rPVI(D) = 100 \times (\sum_{k=1}^{n-1} |d_k - d_{k+1}|)/(n-1)$$

$$nPVI(D) = 100 \times (\sum_{k=1}^{n-1} \frac{|d_k - d_{k+1}|}{(d_i + d_{k+1})/2})/(n-1)$$

A trivial error of interpretation found in the literature is that (*n*-1) excludes final lengthening, but *n*-1 is actually simply the number of differences between adjacent items in a sequence of length *n*. Another often ignored property is that the *rPVI* defines an open-ended linear scale, while the *nPVI* defines a bounded non-linear scale with an asymptote of 200, so the two scales are not commensurable, though they yield the same rankings.

The *PVI* variants are more insightfully seen as minor modifications of standard distance measures than as measures of distributional dispersion. The *rPVI* is essentially Manhattan Distance (also known as Cityblock Distance or Taxicab Distance) and the *nPVI* is essentially Canberra Distance (normalised Manhattan Distance), each measured between two non-disjoint shifted subvectors of *D*:

$$DA = (d_1, …, d_{n-1}), DB = (d_2, …, d_n).$$

The measure has some drawbacks as a measure of rhythm (Gibbon 2003 and many later studies). An empirical issue has been that, although they are often termed 'rhythm metrics', in fact each *PVI* variant removes all rhythmic alternations by taking the absolute value of the subtraction. The *PVI* variants measure evenness of duration and are thus not rhythm metrics per se but *isochrony metrics*. To illustrate: it is easily verified that for both alternating linear 'rhythmic' and non-alternating geometrical vectors, (2,4,2,4,2,4) and (2,4,8,16,32,64), respectively, *nPVI* = 67. Similarly, for both alternating linear (2,4,2,4,2,4) and non-alternating linear (2,4,6,8,10,12) vectors, *rPVI*=200. The same applies to vectors in which alternating and non-alternating subvectors are mixed. Another empirical issue is that the measures are binary (due to the subtraction operation, also shown by interpretation as a distance measure), whereas rhythms may be unary, ternary or more complex, patterns which are beyond the capability of the *PVI* isochrony metrics (cf. also Kohler 2009).

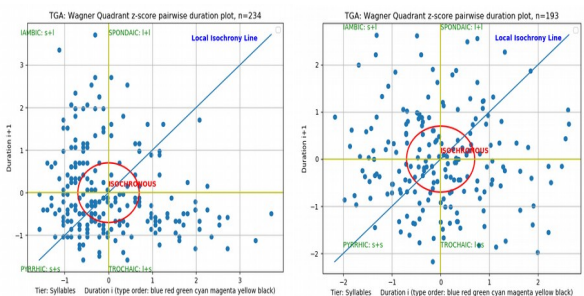

Figure 1: Z-score normalised Wagner scatter plot quadrants, illustrating clear skewness of duration distribution between shorter-shorter syllable pairs (bottom left quadrant) and other pairs for English (left), and more even duration distributions, i.e. more isochronous durations, for Farsi (reading aloud speech styles).

Wagner (2007) has used the binarity of the measure to show the relation between the shifted vectors in a scatter plot, showing two-dimensional details of the distribution of the relation, rather than a simple one-dimensional index.

To summarise: as a model, the PVI is neither empirically 'complete', since it defines a binary relation while rhythms may be more complex (Tilsen and Arvaniti 2013), nor empirically 'sound', since it also measures non-rhythms. The *PVI* variants have nevertheless been very useful as initial heuristics for distinguishing different language types, though was possible essentially because the selected data samples are 'well-behaved' as predominantly alternating and binary, not because the measures inherently distinguish rhythmic from non-rhythmic utterances. Attempts to refute these criticisms (Nolan and Jeon 2014) have not been convincing, which does not detract from the value of the measures as part of a useful 'first response' heuristic, however.

### 2.2. Induction: long-term spectrum rhythm analysis

Parallel to the development of the deductive isochrony models, complementary inductive methods were developed which, in contrast to annotation-based deductive approaches, start from the speech signal and model rhythms as oscillating modulations of the amplitude of the speech signal without reference to linguistic categories. In a posterior step, the results are related to annotations of speech sounds, syllables, words and larger units associated with the speech signal.

Two directions in the inductive, signal-oriented approach developed in parallel: theories of rhythm in speech production and of rhythm in speech perception. The production theories postulate a carrier signal (the fundamental frequency, produced by the larynx) with regularly oscillating amplitude modulations of different frequencies imposed by the changing filter functions of vocal tract shapes (Cummins and Port 1998; O'Dell and Nieminen 1999; Barbosa 2002; Inden et al. 2012). An appropriate spectrum based procedure for modelling rhythm composition in speech production is Fourier synthesis.

Related models of speech perception were independently developed, with demodulation of the amplitude modulations of speech signals, and application of transformations (e.g. Fourier Transform, Hilbert Transform, Wavelet Transform) to extract LF spectral frequencies below about 20 Hz as models of rhythms. The transforms of long sections of the speech signal (approximately >3 s) yield a long-term spectrum (LTS) of the speech signal. In order to obtain this spectrum, the amplitude envelope or intensity trace of the rectified (i.e. absolute) signal (see Section 3) is obtained (Dogil and Braun 1988). Transforms are then applied (Todd and Brown 1994; Cummins et al. 1999; Tilsen and Johnson 2008; Liss et al. 2010; ; Ludusan et al. 2011; Leong et al. 2014; Varnet et al. 2017; Ojeda et al. 2017; He and Dellwo 2016; Gibbon 2018; Šimko et al. 2019; Wayland et al. in press). These studies use slightly different methods, from the intensity trace of Dogil and Braun through the spectra of different frequency bands of Todd and Brown or Tilsen and Johnson to the low spectral frequency *RFT* approach of Gibbon.

Studies of amplitude demodulation with spectral analysis have tended to use elicited data and to focus on higher frequencies in the LTS as indicators of voice quality, for example in clinical phonetic diagnosis, and on lower frequencies as indicators of phonetic rhythm typology. The aim of the present study, in contrast, is to apply *RFT* and novel rhythm demodulation methods to the exploration of formant-like LTS patterns in unelicited discourse, and to include the new dimension of FM (F0, 'pitch') spectral analysis.

### 2.3. Rhythm Formant Theory

*Rhythm Formant Theory* (*RFT*, also: *Rhythm Zone Theory, RZT*) is a further development of the inductive LTS analysis approach. *RFT* makes the following assertions:

1. Modulation. Speech rhythms are low frequency amplitude and frequency modulations of speech, with variable higher magnitude formant-like frequency zones (R-formants), and are tendentially *a fortiori* isochronous.

2. Simultaneous R-formants. The oscillation frequencies of speech rhythms occur in overlapping R-formant frequency ranges, related to units of speech from discourse to phone.
3. Serial R-formants. The R-formants vary with time during discourse, with tempo dependent shifting frequency ranges due to changing speech rates.
4. Asymmetrical rhythm. A tentative novel correlation method is postulated in order to distinguish between physical correlates of abstract strong-weak and weak-strong rhythm patterns, as potential correlates of abstract strong-weak and weak-strong 'metrical' patterns.

*RFT* shares point 1 above with previous LTS-based studies. Points 2, 3 and 4 are innovations. The present study deals with point 2, leaving points 3 and 4 for future investigation. Other differences from previous LTS approaches are:
1. Previous LTS-based approaches (e.g. Todd and Brown 1994, Tilsen and Johnson 1998, Wayland et al. in press) have filtered selected frequency bands out of the signal and analysed these separately. *RFT* takes the spectrum of the entire signal and concentrates on the resulting spectral range below 20 Hz, in particular between 1 Hz and 10 Hz.
2. In the *RFT* approach, LTS frequencies are related explicitly to the inverse average durations of linguistic categories (e.g. syllable, foot) in time-stamped annotations, rather than used *per se*, for example for medical diagnostic purposes, as independent measures of general spectral properties as in earlier approaches.

The main aim of the present study is to continue exploratory development of *RFT* as a new methodology, rather than confirmatory testing of a known methodology on new data. Confirmatory aspects are included, secondarily, in a study of the rhythm of non-elicited public speaking. A tertiary aim is to study how the AMS (LTS of the absolute AM signal), the AEMS (LTS of the AEM), and the FEMS (LTS of the FM/F0/'pitch') are related to each other. The null hypothesis is that the R-formants of the three phonetic domains of AMS, AEMS and FEMS should correlate significantly, as noted in Section 1. If there is no significant correlation, this may support the hypothesis that the AM and FM rhythms have overlapping or independent functions.

Inductive *RFT* analysis is complementary to deductive isochrony measures: linguistic interpretation of *RFT* results requires comparison with the same consonantal, vocalic, syllabic, etc., annotations on which isochrony measures are based, and are thus subject to the same errors of manual or automatic annotation. The isochrony measures themselves are thus complementary to *RFT*, in that the compactness of the formant patterns at particular frequencies relates to isochrony indices for particular linguistic categories such as consonantal, vocalic, syllabic, etc., units. For example, concentration around 4 Hz may relate to low syllabic *nPVI*, and distribution between 3 Hz and 10 Hz or higher may relate to high syllabic *nPVI*. The exact relation is a matter for future investigation.

## 3. Data and Methods

### 3.1. Data

For exploratory purposes, the data set consists of nine orthographically transcribed audio clips taken from video recordings of 2016 USA presidential election campaign speeches by D. J. Trump, obtained from various sources in the internet public domain. The clips are part of a data set of 10 clips, which was originally collected for a study of politeness and impoliteness in public speaking[1] (Li 2018). The first 5 seconds of each clip were extracted in order to ensure comparablity of duration. Utterance 4 from the original selection is excluded from the analysis because it has a duration of only 1 s. Additionally, calibration data consisting of deliberately rhythmical utterances (rapid counting) were created in order to test and demonstrate the validity of *RFT* with known phrasal and syllabic rhythms.

### 3.2. Methods

The study uses *RFT* to investigate relations between R-formants of LF oscillations in the AMS of the rectified signal, the positive envelope, AEM, of the signal, and the FM (F0, 'pitch') of the speech signal. The positive spectra in the three domains are log normalised, and residuals of a linear regression function are extracted in order to create a near-flat baseline for magnitude measurements. For visualisation, the result is squared in order to amplify differences. LF segments from 1 to 10 Hz are extracted from the normalised transforms. Vectors with a specifiable number $n$ of frequencies (here: $n$=6) with the highest magnitudes are obtained from the LF segments, and 10 histogram bins weighted with the magnitudes of these frequencies, representing the magnitudes of R-formants, are formed from the vectors.

Pearson's $r$ between the AMS, AEMS and FEMS spectra is obtained pairwise from the bin values. The bins are also examined with the Mantel permutation test in order to check whether any correlation at all exists globally (without preserving sequential order) between the R-formants in the three domains. The non-autocorrelation condition for the Mantel test is fulfilled.

The bins for each domain are collated in a 9 × $n$ matrix, and Manhattan distances between the domains are calculated pairwise for each utterance. Additionally, but not part of the comparison procedure, the relation between the R-formants of the items in the data set was visualised as a dendrogram, using Manhattan distance and average pair group linkage.

The analyses are implemented in Python (compatible with versions 2.7.n and 3.n), with libraries NumPy, MatPlotLib, SciPy, and a Mantel permutation test module. An extensive set of display and analysis parameters can be set in a configuration file.[2] An online version is also available.[3]

### 3.3. Illustration of the Rhythm Formant analysis method

The validity of the R-formant analysis method is illustrated with the rhythmically clear case of regular fast counting in English (Figure 2).

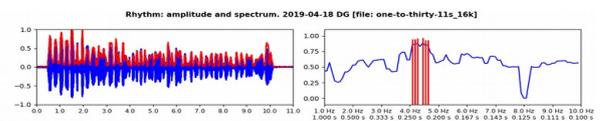

Figure 2: Acoustic analysis of prosody parameters for a rhythmical test utterance (rapid counting from 1 to 30).

The visualisation contains the following information:
1. Left: waveform with positive amplitude envelope outlining the upper, positive half of the waveform;
2. Right: low frequency long-term spectrum (AMS, i.e. LF LTS) of the absolute (rectified) waveform, with 'rhythm bars' (vertical lines in the spectrum) showing a formant-like cluster (R-formant) of 6 spectral frequency components with the greatest magnitude.

Lexically, the numerals are monosyllables, disyllables, trisyllables and quadrisyllables ('27'), but the fast speech rendering results in some weak syllable deletions, e.g. in '7'.

Analysis of an annotation of the utterance shows:
1. durations of intervals tend to increase during the utterance;

---

1. Online survey form with downloadable audio clip data: http://wwwhomes.uni-bielefeld.de/gibbon/OSCAR/OSCAR_al02/
2. The code is freely available under the GNU General Public License v3.0 licence at https://github.com/dafyddg
3. http://wwwhomes.uni-bielefeld.de/gibbon/CRAFT

2. the *nPVI* shows regular durations: syllable *nPVI* = 44 (read-aloud English: around 60), word *nPVI* = 11, strong syllable *nPVI* = 25;
3. Word count is 30, total duration 9.667 s, mean duration 322 ms, and fast word rate is 3.1/s (3.1 Hz). Mean syllable duration is 161 ms, and fast mean syllable rate 6.21/s (6.21 Hz).

The prediction based on the annotated values is that dominant frequencies in the AMS will range from about 3.1 Hz to 6.21 Hz, with a centre frequency around 4.7 Hz.

Figure 2 shows the region 1 Hz to 10 Hz of the AMS, with a cluster of 6 dominant frequencies between about 4.1 Hz and 4.6 Hz. The mid frequency of about 4.35 Hz differs from the predicted approximation of 4.7 Hz by only 0.35 Hz, an informal corroboration of the prediction.

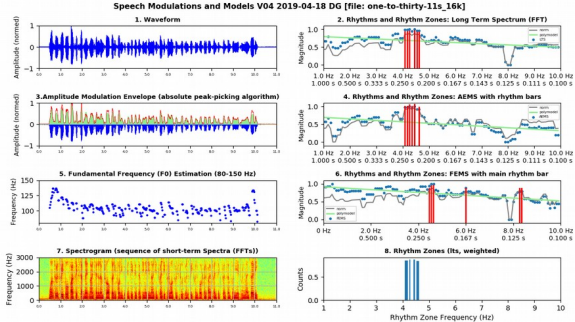

Figure 3: Acoustic signal analysis of prosody parameters for a deliberately rhythmical test utterance (rapid counting from 1 to 30).

Additional measurements were performed in order to determine other relevant prosodic properties (Figure 3):
1. top left: waveform;
2. top right: low frequency long-term spectrum (AMS) of the rectified absolute signal, with rhythm bars (vertical red lines in the spectra), showing frequencies with the greatest magnitudes;
3. upper mid left: waveform with the amplitude envelope outline of the rectified (absolute) signal, made with a peak-picking moving window (some approaches use the absolute Hilbert Transform);
4. upper mid right: low frequency long-term spectrum of the amplitude envelope modulation (AEMS) of the rectified (absolute) signal, with rhythm bars;
5. lower mid left: frequency modulation (FM, F0 estimation, 'pitch tracking') using a custom implementation of the Average Magnitude Difference Function (AMDF) algorithm;
6. lower mid right: LF FM spectrum (FEMS);
7. bottom left: spectrogram;
8. bottom right: histogram of the AMS, clustering frequencies weighted by their magnitudes to visualise R-formants.

The top right and upper mid right graphs are interpreted as showing similar R-formants within the segment 1 Hz to 10 Hz of the long-term AMS of the amplitude variations in the signal, though with different analytic procedures.

The lower mid right shows the FEMS, which does not correspond to the AMS and AEMS, presumably because of independent variation in intonation and pitch accent contours.

The bottom right graph in Figure 3 shows R-formants, reflecting AMS rhythm bar patterns.

Based on general phonetic background knowledge, it may be assumed that frequencies above about 10 Hz relate to syllable components and shorter syllables of lengths 100 ms and below, frequencies around 4 Hz relate to longer syllables of lengths around 250 ms, while words and foot-length units of about 700 ms are related to frequencies of around 1.5 Hz. Frequencies below 1 Hz relate to rhythmic regularities of phrases and larger discourse units which are longer than 1 s.

These values, like the isochrony measures, are dependent on the speaker and on whether speech style is fast or slow, and may vary from language to language or dialect to dialect.

## 4. Results

### 4.1. Initial description and comparisons

Figures 4 and 5 illustrate the corpus analysis using the first 5 s of audio clips 2 and 9 from the corpus. Clip 2 is in narrative style, while clip 9 is emphatic with rhetorical pauses between lexical groups. The leftmost graph in each figure shows waveform and amplitude envelope, while the rightmost graph shows the amplitude spectrum derived by Fourier Transform, and rhythm bars for the 16 most prominent frequencies.

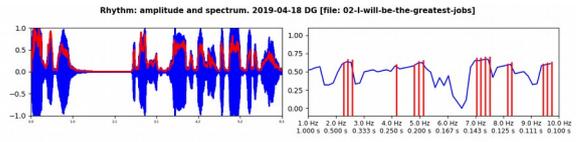

Figure 4: Acoustic analysis of survey utterance 2: *I will be – the greatest jobs president – that God ever created.* Dashes mark pauses.

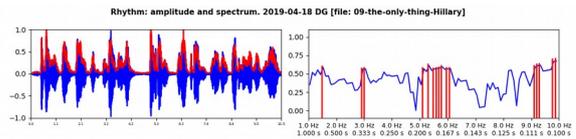

Figure 5: Acoustic analysis of survey utterance 9: *The only – thing – that Hillary – Clinton – has going – for herself – is the press. – Without the press – she is absolutely – zero.* Dashes mark pauses.

The R-formants are distributed very differently in these examples, and both differ from the counting example (Figures 2 and 3). The narrative style utterance in Figure 4 shows relatively diverse rhythmic variation in rather slow speech, with R-formants around 2.5 Hz (400 ms, foot size), 5 Hz (200 ms, approximately strong syllable size) and 7.25 Hz (130ms, approximately weak syllable size). The rhetorically emphatic utterance in Figure 5, with even slower speech, shows a concentration of rhythmic beats in a dominant R-formant centring on about 6 Hz (units averaging 170 ms).

### 4.2. AMS, AEMS and FEMS domain comparisons

The question arises of how the AMS, AEMS and FEMS domains are related to each other, the null hypothesis being that the spectra in the three domains correlate significantly. In order to test the global correlation between the domains, the strict Mantel permutation test (which yields the best Pearson correlation among all permutations of the data under comparison) was employed, even overriding the sequentiality of the data, with a clear result (Table 1).

Table 1: Global Mantel permutation test results for distance matrices.

|  | *r* | *p* | *significance* |
|---|---|---|---|
| **AEMS:FEMS** | -0.38 | 0.1 | ns |
| **AMS:AEMS** | 0.47 | 0.01 | ** |
| **AMS:FEMS** | -0.11 | 0.6 | ns |

The significant, though not strong AMS:AEMS correlation confirms the null hypothesis. The amplitude and frequency spectrum correlations have no significant correlation, thus refuting the null hypothesis of similarity between the spectra.

This result suggests that the amplitude and frequency spectral domains are partially independent. The use of other F0 tracking algorithms may also contribute to clarifying the result (Gibbon 2019). However, the Mantel permutation test may also simply be too global in view of the sequentially ordered data items.

To gain more detailed information, pairwise correlations between the AMS, AEMS and FEMS domains for each of the audio clips were obtained (Table 2).

Table 2: Correlation (Pearson's *r*) means for AEMS:FEMS, AMS:AEMS and AMS:FEMS spectrum pairs over all utterances (mean *r*). Utterances with the smallest (min *r*) and largest (max *r*) correlations for each spectrum pair are also listed.

|  | mean *r* | Utt | min *r* | Utt | max *r* |
|---|---|---|---|---|---|
| **AEMS:FEMS** | 0.07 | 6 | -0.5 | 8 | 0.67 |
| **AMS:AEMS** | 0.6 | 1 | -0.07 | 7 | 0.96 |
| **AMS:FEMS** | 0.09 | 6 | -0.4 | 8 | 0.76 |

Table 2 shows that on average amplitude spectra AMS and AEMS correlate, while amplitude and frequency spectra, AMS and AEMS on the one hand and FEMS on the other, do not, with exceptions. Further investigation on variation of tone, pitch accent and intonation contours is needed in order to explain this difference.

### 4.3. Classification of R-formants

A closer look at the histogram patterns suggests that utterances may be classifiable on the basis of similarity in holistic rhythm gestalt. In the following analysis, only the AMS is used.

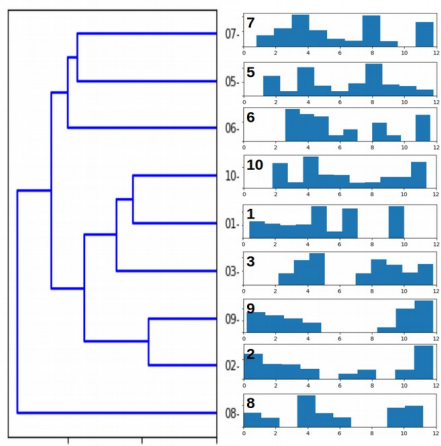

Figure 6: Classification of utterance distance matrices with Manhattan distance and Average Pair Group Method with Arithmetic Mean (UPGMA) linkage, with superimposed R-formant histograms.

AMS frequency bins for the nine audio clips (cf. Section 8, Data Appendix), weighted by frequency magnitudes, are shown in Figure 6, with Manhattan distance classification of the long-term spectra of the utterances (utterance 4 is very short and was excluded). The dendrogram shows a similarity hierarchy of the AMS spectral bins for the survey audio clips. Visual inspection confirms obvious similarities.

The spectrum segments represented in the bins of Figure 6 are between 0 Hz and 12 Hz, with prominent LF concentrations at around 0.5...1 Hz, which are characteristic of rhetorical regularity in phrasal segments. As noted in Section 3.3, R-formants around 3 Hz to 5 Hz are characteristic of longer, e.g. stressed syllables, while the R-formants above about 7 Hz are characteristic of shorter, e.g. unstressed syllables and of syllable constituents, with variation in R-formant positions depending on speech tempo.

### 4.4. Tentative interpretation with pragmatic variables

A first informal hypothesis for a pragmatic interpretation of the classification is that lower frequency clusters below 3 Hz indicate rhetorical emphasis and pausing, while higher frequency clusters above 3 Hz indicate a more neutral narrative style (cf. Li 2018).

Based on text-grammatical properties of the audio clips a grouping into two classes emerges, corresponding to the two main classes in Figure 6 (see Section 8, Data Selection): clips with a more verbose narrative style (1, 2, 3, 9, 10) and clips with a more concise exhortative style (5, 6, 7, also 8).

However, the original motivation for the data collection was an opinion survey on politeness descriptors for public speaking (Li 2018); cf. Figure 7.

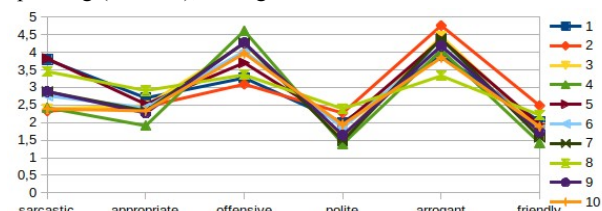

Figure 7: Attribute rating means for all raters (*y*-axis) of 5-point Likert attributions (*x*-axis) for each of 10 Trump campaign speech clips.

It is immediately obvious by visual inspection that all 10 utterances received very similar rating distributions. While this was useful for the original study, the prediction for present purposes has to be that it is unlikely that a useful degree of correlation of these similar patterns with the diverse R-formant patterns of the *RFT* analysis can be found. As expected, the Mantel permutation test yields no correlation for most distance measures, but for Hamming distance it unexpectedly yields a weak correlation of 0.24 between the opinion score results and the FEMS results for distance matrices, a sign that a study with more differentiated pragmatic categories could be useful.

Further *post hoc* pragmatic interpretation of the audio clips showed a common interpretation of the group 01-10-03 as dissatisfaction and denunciation of others, for the 02-09 group self-praise and deriding of others as positive and negative extremes, for 05-06-07 belittling and mocking of others, and for 08, an outlier, a superficially factual utterance.

## 5. Summary, conclusions and outlook

The present study applies a new form of rhythm analysis, *Rhythm Formant Theory* (*RFT*) involving long-term spectrum (LTS) analysis, identification of higher magnitude spectral frequency zones (rhythm formants, R-formants) in the LTS, and classification of utterances by R-formant patterns found in a small corpus of non-elicited public speech data.

After an overview of complementary deductive and inductive approaches to the study of the rhythms of speech, the empirical viability of *RFT* analysis was first grounded in a clear case of R-formants in obviously rhythmical data: rapid counting. Then *RFT* analysis of three prosodic acoustic spectral domains was undertaken with the corpus, analysing the low frequency segments of the long-term spectra of the rectified (absolute) amplitude modulation (AM) of the speech signal, the positive envelope of the amplitude modulation of the speech signal and the frequency modulation (FM, F0, 'pitch') of the speech signal. The AM spectral domains correlate significantly, but the FM spectral domain does not correlate consistently with the AM spectral domains, suggesting relative structural and functional independence of the AM and FM domains. R-formant patterns of the corpus utterances were classified hierarchically and related to pragmatic categories.

Areas for future study and for development of R-formant analysis as a potential rhythm research paradigm were pointed out. Much further research using complementary annotation-based approaches is needed in order to be able to interpret *RFT* results structurally and functionally, by comparing annotations of phonological and prosodic units with R-formants, and by relating compactness of R-formant patterns with isochrony indices. Results for pragmatic categories are tentative, but are useful as pointers to further research topics.

The selection of data for the present exploratory study was limited by design, and, clearly, more extensive data resources and more detailed investigation of appropriate algorithms, as well as of pragmatic descriptors, are needed in order to be able to achieve more generally valid results.

# 6. Acknowledgments

The authors are grateful for critical feedback from the anonymous reviewers and, at various times, by Laura Dilley, Alexandra Gibbon, Xuewei Lin, Rosemarie Tracy, Ratree Wayland and Jue Yu. Special thanks are due to Huangmei Liu for suggesting and motivating the term *formant* for *zone*.

# 8. Data Appendix

The survey with transcripts and audio clips are available at: http://www.uni-bielefeld.de/gibbon/OSCAR/OSCAR_al02/

The transcripts of the the 10 audio clips, numbered as referred to in the survey and in the paper, are as follows:

1. They met for thirty-nine minutes, remember, he said: "We talked golf, and we talked about our grandchildren." Three minutes for the grandchildren, two minutes for the golf.
2. There has never been in the history of the world a greater theft than what China did to the US. We have rebuilt China, they have taken our jobs, they have taken our base, they have taken our money, they have taken everything, they have drained us.
3. Our country is being killed, because we have stupid people leading our country. We have people that don't know what they are doing! They don't know what they're doing!
4. [ Too short, not used. ]
5. I will be the greatest jobs president that God ever created.
6. The only thing Hillary Clinton has going for herself is the press. Without the press, she is absolutely zero.
7. Five billion dollar website, I have so many websites, I have them all over the place. I hire people, they do a website, it costs me three dollars. Five billion dollar website.
8. How stupid are our leaders! How stupid are these politicians to allow this to happen! How stupid are they.
9. And I, I have to be honest with you. I don't think Hillary has the strength or the stamina to be president. I really mean that.
10. Then there are the thirty-three thousand emails she deleted. While we may not know what's in those deleted emails, our enemies probably know every single one of them.